\documentclass[sigplan,10pt]{acmart}
\renewcommand\footnotetextcopyrightpermission[1]{}
\setcopyright{none}
\AtBeginDocument{%
  \providecommand\BibTeX{{%
    \normalfont B\kern-0.5em{\scshape i\kern-0.25em b}\kern-0.8em\TeX}}}

\usepackage{xspace}
\usepackage{multirow}
\usepackage{comment}
\usepackage{xfrac}
\usepackage{fancybox, fancyvrb, calc}
\usepackage[center, tight]{subfigure}
\usepackage[inline]{enumitem}
\usepackage{amsmath}
\usepackage{blkarray, bigstrut}
\usepackage[font=small]{caption}
\usepackage{url}
\usepackage{hyperref}
\usepackage{graphicx}
\usepackage{pbox}
\usepackage{algorithm, algorithmicx}
\usepackage[noend]{algpseudocode}
\usepackage{balance}
\usepackage{wrapfig}
\usepackage{tikz, pgfplots}
\usetikzlibrary{patterns,matrix,positioning,shadows,backgrounds,arrows,calc,fit,automata,shapes.geometric,arrows,decorations.pathreplacing,decorations.markings,shadings,shapes.symbols,arrows.meta}
\usepackage[capitalise]{cleveref}
\usepackage{tabularx}
\usepackage[T1]{fontenc}
\usepackage[utf8]{inputenc}
\usepackage[normalem]{ulem}
\usepackage{wasysym}
\usepackage{enumitem}
\usepackage{sidecap}
\usepackage{caption}
\usepackage{pifont}
\usepackage[mathcal]{euscript}
\usepackage{setspace}
\usepackage{listings}
\usepackage{xcolor}
\usepackage{footnote}
\usepackage{times}
\makesavenoteenv{tabular}
\makesavenoteenv{table}
\usepackage[noindentafter]{titlesec}
\usepackage{tikz}
\usepackage{titlesec}   

\titlespacing\section{0pt}{3pt plus 1pt minus 1pt}{3pt plus 1pt minus 1pt}
\titlespacing\subsection{0pt}{3pt plus 1pt minus 1pt}{3pt plus 1pt minus 1pt}

\usepackage{hyperref}
\hypersetup{
  colorlinks=true,      
  linkcolor=blue,       
  citecolor=magenta,    
  filecolor=cyan,       
  urlcolor=red          
}
\definecolor{listinggray}{gray}{0.9}
\definecolor{lbcolor}{rgb}{0.9,0.9,0.9}
\graphicspath{{fig/}}

\usepackage{etoolbox}

\definecolor{boxclr}{gray}{0.9}

\pgfdeclarelayer{bg}    
\pgfsetlayers{bg,main}  

\makeatletter
\newcommand{\thickhline}{%
    \noalign {\ifnum 0=`}\fi \hrule height 0.8pt
    \futurelet \reserved@a \@xhline
}
\newcolumntype{"}{@{\vrule width 0.8pt}}
\newcolumntype{[}{@{\vrule width 0.8pt\hskip\tabcolsep}}
\newcolumntype{]}{@{\hskip\tabcolsep\vrule width 0.8pt}}
\newcolumntype{!}{@{\hskip\tabcolsep\vrule width 0.8pt\hskip\tabcolsep}}
\makeatother

\newcommand{\cppsnippet}[1]{%
  \begin{lstlisting}[gobble=4]
    #1
  \end{lstlisting}
}

\tikzset{
    position/.style args={#1:#2 from #3}{
        at=(#3.#1), anchor=#1+180, shift=(#1:#2)
    }
}

\tikzset{
  half fill/.style 2 args={fill=#2, path picture={
    \fill[#1, sharp corners] (path picture bounding box.west) --
                         (path picture bounding box.east) --
                         (path picture bounding box.south east) --
                         (path picture bounding box.south west) -- cycle;}},
}

\tikzset{
  nil fill/.style 2 args={fill=#2, path picture={
    \fill[#1, sharp corners] (path picture bounding box.155) --
                         (path picture bounding box.25) --
                         (path picture bounding box.south east) --
                         (path picture bounding box.south west) -- cycle;}},
}

\tikzset{
  almost fill/.style 2 args={fill=#2, path picture={
    \fill[#1, sharp corners] (path picture bounding box.205) --
                         (path picture bounding box.335) --
                         (path picture bounding box.south east) --
                         (path picture bounding box.south west) -- cycle;}},
}

\definecolor{oldcolor}{HTML}{c66541}

\newcommand*\circled[1]{\tikz[baseline=(char.base)]{
            \node[shape=circle,draw,inner sep=0.5pt] (char) {\small #1};}}

\algdef{SE}[DOWHILE]{Do}{doWhile}{\algorithmicdo}[1]{\algorithmicwhile\ #1}%

\newcommand{\cut}[1]{}

\newcommand{\paragraphb}[1]{\vspace{0.075in}\noindent{\bf #1.}}

\newcommand{\paragraphc}[1]{\vspace{0.075in}\noindent{\em #1}}

\tikzset{ 
table/.style={
  matrix of nodes,
  row sep=-\pgflinewidth,
  column sep=-\pgflinewidth,
  nodes={rectangle,thick,draw=black,text width={},align=center,font=\small},
  text depth=0.25ex,
  text height=1.25ex,
  nodes in empty cells
},
map/.style={
  matrix of nodes,
  row sep=-\pgflinewidth,
  column sep=-\pgflinewidth,
  nodes={rectangle,draw=black,text width=5em,align=center,font=\small},
  text depth=0.25ex,
  text height=1.25ex,
  nodes in empty cells
},
bigmap/.style={
  matrix of nodes,
  row sep=-\pgflinewidth,
  column sep=-\pgflinewidth,
  nodes={rectangle,draw=black,text width=26em,align=center,font=\small},
  text depth=0.25ex,
  text height=1.25ex,
  nodes in empty cells
},
memcell/.style={
  draw, 
  very thick, 
  text width=0.25em, 
  text height=0.25em
},
}

\tikzstyle{startstop} = [rectangle, rounded corners, minimum width=3em, minimum height=1em,text centered, draw=black, fill=red!30]
\tikzstyle{io} = [trapezium, trapezium left angle=70, trapezium right angle=120, minimum width=2.5em, minimum height=1em, text centered, draw=black, fill=blue!30]
\tikzstyle{process} = [rectangle, minimum width=1.5em, minimum height=1em, align=center, draw=black, fill=gray!30]
\tikzstyle{decision} = [diamond, minimum width=3em, minimum height=1em, align=center, draw=black, fill=SkyBlue!30]
\tikzstyle{arrow} = [thick,->,>=stealth]
\tikzstyle{monolog} = [fill=SkyBlue!30]

\tikzset{
  on each segment/.style={
    decorate,
    decoration={
      show path construction,
      moveto code={},
      lineto code={
        \path [#1]
        (\tikzinputsegmentfirst) -- (\tikzinputsegmentlast);
      },
      curveto code={
        \path [#1] (\tikzinputsegmentfirst)
        .. controls
        (\tikzinputsegmentsupporta) and (\tikzinputsegmentsupportb)
        ..
        (\tikzinputsegmentlast);
      },
      closepath code={
        \path [#1]
        (\tikzinputsegmentfirst) -- (\tikzinputsegmentlast);
      },
    },
  },
  mid arrow/.style={postaction={decorate,decoration={
        markings,
        mark=at position .5 with {\arrow[#1]{stealth}}
      }}},
}

\newcommand{\specialcell}[2][l]{%
  \begin{tabular}[#1]{@{}l@{}}#2\end{tabular}}

\algnewcommand{\IIf}[1]{\State\algorithmicif\ #1\ \algorithmicthen}%
\algnewcommand{\EndIIf}{\unskip\ }%
\algdef{SE}[DOWHILE]{Do}{doWhile}{\algorithmicdo}[1]{\algorithmicwhile\ #1}%
\algnewcommand\algorithmicforeach{\textbf{for each}}%
\algdef{S}[FOR]{ForEach}[1]{\algorithmicforeach\ #1\ \algorithmicdo}%

\newcommand{\code}[1]{{\fontsize{9}{11}\selectfont\texttt{#1}}}
\newcommand{\scode}[1]{{\fontsize{7}{11}\selectfont\texttt{#1}}}

\def\ie{{i.e.}}
\def\eg{{\em e.g.}\xspace}

\def\etc{etc.}

\def\designnamex{{GCP}}
\def\designname{\designnamex\xspace}
\def\namex{{Soul}}
\def\name{\namex\xspace}

\usepackage{soul}
\setulcolor{red}
\newcommand{\hlc}[2][yellow]{ {\sethlcolor{#1} \hl{#2}} }

\colorlet{soulblue}{blue!20}

\colorlet{soulpink}{pink!40}
\newcommand\yanpengdelete[1]{\hlc[soulpink]{}}
\colorlet{soulgreen}{green!30}

\colorlet{soulred}{blue!20}

\colorlet{soulorange}{orange!20}

\definecolor{codegreen}{rgb}{0,0.6,0}
\definecolor{codegray}{rgb}{0.5,0.5,0.5}
\definecolor{codepurple}{rgb}{0.58,0,0.82}
\definecolor{backcolour}{rgb}{0.95,0.95,0.92}

\begin{document}\sloppy
\title{\name: Generalized Cache Coherence For Efficient Synchronization}

\author{Yanpeng Yu\qquad Seung-seob Lee\qquad Anurag Khandelwal\qquad Lin Zhong\\Yale University}
\begin{abstract}

We explore the design of scalable synchronization for disaggregated shared memory. Porting existing synchronization primitives to such memory results in poor performance scaling with the number of application threads because these primitives are layered atop cache-coherence substrates. This layering engenders redundant inter-cache communications,  which are exacerbated by the high cache-coherence latency ($\mu$s) with low bandwidths in state-of-the-art disaggregated shared memory designs, precluding application scalability.

In this work, we argue for a co-design of the cache-coherence and synchronization layers for better performance scaling of multi-threaded applications on disaggregated memory. This is driven by our observation that synchronization is essentially a generalization of cache coherence in time and space. We present \name as an implementation of this co-design. \name employs wait queues and arbitrarily-sized cache lines directly at the cache-coherence layer for temporal and spatial generalization, respectively. We evaluate \name against state-of-the-art locks and show that \name improves in-memory key-value store and database management system performance at scale by $1-2$ orders of magnitude.

\end{abstract}
\maketitle
\pagestyle{plain}
\pagestyle{plain}

\section{Introduction}

Synchronization primitives such as locks~\cite{nonscalablelocks, litl, everthingsync, mcslock, clhlock, cohortlock, hmcslock, cnalock, shufflelock, cloflock, linuxbrlock, passiverwlock, numarwlock, bravo} are crucial for the performance of multi-threaded applications. In multi-core CPU architectures, synchronization primitives are built atop an efficient hardware-based cache coherent substrate that ensures atomicity for a unit of cache data\footnote{The atomic memory abstraction~\cite{nagarajan2020primer}} (\ie, a cache line), by extending this atomicity guarantee to multiple cache lines spread across an arbitrary period of time. In this work, we focus on locks since they form the core building block for most other synchronization primitives.

With the end of Moore's Law and the consequent challenges in scaling DRAM technologies~\cite{memscaling1} in a single server, recent years have seen a push towards rack-scale compute-memory disaggregation~\cite{industry0, industry1, industry2, industry3, industry4, memdisagg1, memdisagg2, memdisagg3, memdisagg4, memdisagg5, memdisagg6, infiniswap, fastswap, legoos, mind}, where server resources are physically decoupled into compute and memory blades connected via a high-speed network fabric, with the compute blades equipped with a small amount of DRAM as a cache. Recent efforts have also focused on enabling cache-coherent shared memory abstractions over them for application transparency~\cite{mind, cxl}. It stands to reason that the lock algorithms developed for multi-core architectures could be ported to cache-coherent substrates atop disaggregated memory.

Unfortunately, the high latency and low bandwidth of the inter-cache connections in disaggregated memory make its cache-coherent substrates inadequate for an efficient realization of lock algorithms. In particular, while inter-cache communications in multi-core and NUMA architectures observe latencies around $20$--$100$ ns and operate over a bandwidth upwards of $500$ Gbps, those for disaggregated memory see latencies of $5$--$10~\mu$s while the bandwidth drops to $100$ Gbps, even with RDMA~\cite{mind, gam, concordia}. As such, even the most high-performance lock algorithms layered over state-of-the-art coherent substrates for disaggregated memory~\cite{mind, cxl} observe poor performance scaling. 


Prior software-based distributed shared memory (DSM) systems~\cite{1994kelehertWTC-treadmarks, munin, midway, gam} face a similar challenge. They choose to \emph{bypass} the cache-coherence layer and build optimized lock services that leverage weaker memory consistency models like PSO~\cite{gam} or release consistency~\cite{1992keleherISCA-releaseconsistency}. 
Two problems arise when adapting such services for disaggregated shared memory, leading to sub-optimal performance and application complexity.
First, they require applications to interact with two distinct services --- cache-coherent shared memory and lock --- requiring careful modification of application logic to ensure correctness. 
Second, they cannot benefit from the efficient hardware-based cache-coherent interconnects featured in state-of-the-art disaggregated shared memory systems~\cite{mind, cxl}. 
While some recent distributed systems have explored hardware-based realization of their lock services~\cite{lockservice2, lockservice3, netlock}, they require additional hardware complexity in an already resource-constrained inter-compute interconnect (\S\ref{ssec:softwaredsm}).


This raises the question: \emph{Is it possible to design scalable and high-performance locks for disaggregated shared memory
 leveraging 
the existing cache-coherent substrates?} We answer the question in the affirmative with a principled redesign of lock-based synchronization, drawing on two key observations.

First, we find that the key reason behind the poor scalability and performance of shared memory lock algorithms on disaggregated memory lies in the redundant inter-cache communications when lock algorithms are layered atop cache coherence (\S\ref{ssec:layereddesign}). In fact, a closer inspection (\S\ref{sec:design}) reveals a far more direct relationship between lock-based synchronization and cache coherence --- lock-based synchronization is a generalization of cache coherence in time and space:
\begin{itemize}[leftmargin=*, itemsep=0pt]
  \item \textbf{Temporal generalization.} Cache-coherence protocols guarantee the single-writer-multi-reader (SWMR) invariant~\cite{nagarajan2020primer} --- either a single exclusive writer or multiple concurrent readers can access a cache line --- for the duration of a single instruction. In contrast, lock-based synchronization requires this property for an arbitrary number of instructions --- specifically, the critical section.
  \item \textbf{Spatial generalization.} Similarly, cache-coherence protocols ensure the SWMR invariant at a cache line granularity ($64$B in most CPUs), while reader-writer locks typically require this property for shared states of arbitrary sizes.
\end{itemize}
\noindent
This suggests that an extension of existing cache-coherence protocols that supports these temporal and spatial generalizations would provide the necessary semantics to directly serve as lock primitives, and thus eliminate the redundant communications seen in a layered design. Moreover, co-designing synchronization with coherence also enables opportunities for other optimizations, \eg, caching lock and shared data until explicitly invalidated, pipelining movement of shared data (\ie, the cache line in coherence protocols) with lock acquisition, \etc, enabling further improvement in application performance.


Second, while inter-cache communications in disaggregated systems do observe higher latencies and lower bandwidths, making the overheads of the layered design far more pronounced than in traditional multi-core architectures, they are also inherently more flexible due to their programmable cache-coherent interconnects~\cite{mind}. We argue that the same flexibility can enable the temporal and spatial generalization of coherence protocols described above.


We incorporate the above insights into \designname, a novel class of \textit{G}eneralized \textit{C}ache-coherence \textit{P}rotocols for lock-based synchronization. \designname minimally modifies directory-based cache-coherence protocols using two key ideas:
\begin{itemize}[leftmargin=*, itemsep=0pt]
    \item \textbf{Wait queues for temporal generalization.} Cache coherence only guarantees that a cache line requested by a thread will be held in the requestor's cache for a single instruction --- subsequent requests from other threads \emph{invalidate} the cache line on the original requestor. To allow a requestor to hold the cache line for more than one instruction, \designname prevents other requestors from invalidating the cache line until the original requestor explicitly releases it; the other requestors are instead recorded in a wait queue, and their execution is suspended until it is their turn to access the cache line\footnote{The order in which requestors get to access the cache line is a matter of policy that has been extensively studied in prior work, \eg, FIFO, random, priority order, \etc --- \designname does not innovate on this front.}. Note that the use of wait queues for synchronization is not new; however, our approach differs in embedding them at the \emph{cache-coherence layer}.
    \item \textbf{Arbitrarily-sized cache lines for spatial generalization.} Spatial generalization is relatively straightforward --- while directory-based cache-coherence protocols track a single fixed-sized cache line at each directory entry, our generalization simply tracks a list of arbitrarily-sized memory regions per directory entry. During invalidations, all of the shared regions tracked by the directory entry are removed from the target cache. This allows threads to achieve atomic access to arbitrary-sized shared regions for the duration of their critical section. 
\end{itemize}
\vspace{-0.5em}
\noindent
These generalizations essentially permit transforming some directory entries in the cache-coherence protocol into lock entries for synchronizing shared state in \designname. We note, however, that these modifications do not affect the operation of cache coherence for regular cache entries --- they still operate at a single cache line and single instruction granularity. Moreover, since all of our modifications are confined to the cache-coherence protocol, \designname does not impose any restrictions on the memory consistency model, \ie, it can work with any consistency model ranging from more relaxed release consistency models to stricter TSO consistency models.

We present \name as an implementation of \designname atop MIND~\cite{mind}, a state-of-the-art disaggregated shared memory system that employs in-network directory-based MSI cache-coherence protocol with TSO memory consistency. \name places a minimal amount of state and logic associated with coherence in the programmable network switch to work around the switch's limited resources while delegating much of it to the compute blades. \name also employs a novel queue transfer protocol to minimize the network latency for accessing the wait queue associated with a cache line in \designname. 

\name enables complete transparency for legacy shared memory applications by wrapping \designname in a POSIX-compliant reader-writer lock API.
\name also exposes synchronization APIs in modern languages like Rust for programmers to explicitly specify memory regions to be protected by a lock during lock initialization. With this API, since the shared memory regions accessed by a critical section are already tracked by the directory entries, \name proactively combines the movement of corresponding data with lock acquisition to minimize data access latency during the critical section~(\S\ref{ssec:designapi}). Our evaluation shows that \name enables scalable and high-performance lock-based synchronization on disaggregated memory. Compared to state-of-the-art lock algorithms and standalone lock services, \name improves the in-memory key-value store and database management system performance at scale by $1-2$ orders of magnitude.

\section{Motivation}
\label{sec:motivation}





We begin with a brief background on cache coherence and lock-based synchronization(\S\ref{ssec:background}) and then demonstrate the inefficiency of a layered design for the two (\S\ref{ssec:layereddesign}) and standalone lock services of software-based DSM systems ({\S\ref{ssec:softwaredsm}}).

\subsection{Background}
\label{ssec:background}

\paragraphb{Cache coherence} Most modern CPU architectures use either \emph{snoop-based} or \emph{directory-based} cache-coherence protocols.
We focus on directory-based ones in this work because most systems of NUMA-scale or larger (including proposals for disaggregated memory~\cite{cxl, mind}) use them for their scalability.

Directory-based protocols employ a logically centralized cache directory to track the state of the basic memory unit --- typically referred to as a cache line --- present in the distributed caches. The state at the directory includes the list of caches that currently hold the cache line (\ie, the ``sharer list''), and what permissions they have. When any cache intends to acquire a cache line, it first contacts the directory, which subsequently notifies other caches in the sharer list about the transition and coordinates any subsequent data movement.

\begin{figure}[t!]
\centering
    \includegraphics[width=0.3\textwidth]{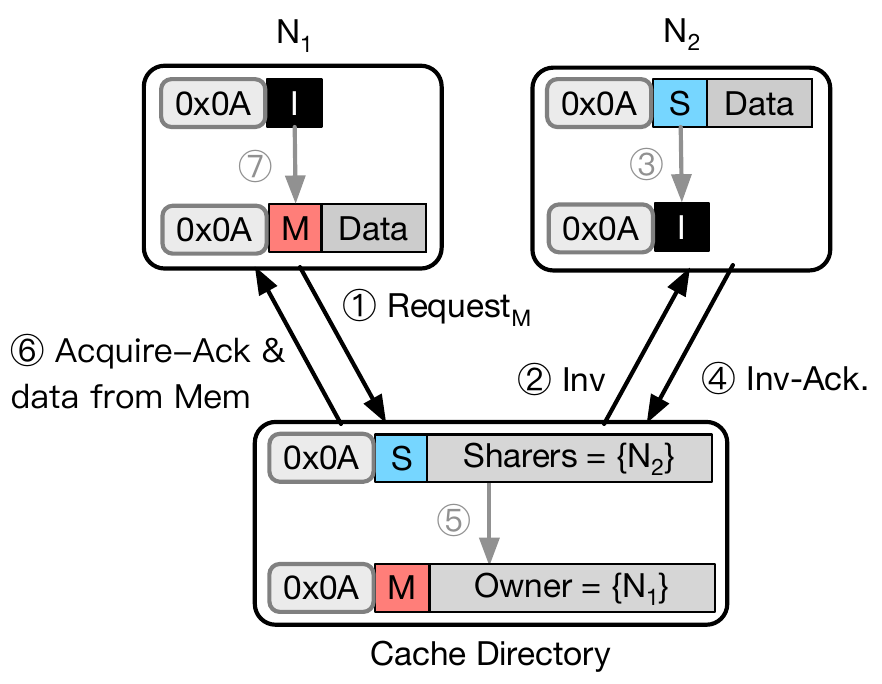}\vspace{-1em}
  \caption{\textbf{Directory-based MSI Protocol (\S\ref{ssec:background})}: a single cache line at address \code{0x0A}, a cache directory, and two nodes --- $N_1$ and $N_2$.}
  \label{fig:cache_coherence_example}
  \vspace{-1.5em}
%
\end{figure}

Consider a simple directory-based MSI protocol, where each cache line can be in one of the three permissions:
\begin{itemize}[itemsep=0pt, leftmargin=*]
  \item \textbf{M} or modified, indicating a single cache has \emph{exclusive} read and write permissions for the cache line,
  \item \textbf{S} or shared, where multiple caches have \emph{shared} read permission to the cache line, and,
  \item \textbf{I} or invalid, \ie, the cache line is not present in any cache.
\end{itemize}
\noindent
Fig.~\ref{fig:cache_coherence_example} shows an example with a single cache line at address \code{0x0A}, a cache directory, and two nodes --- $N_1$ and $N_2$. The cache line is initially cached at $N_2$ with \textbf{S} permission. $N_1$ then requests it with \textbf{M} permissions from the cache directory (\circled{1}). The directory looks up the sharer list for the cache line and contacts $N_2$, its current holder. Specifically, the cache directory must \emph{invalidate} the cache line at $N_2$, since $N_1$ needs exclusive access to it (\circled{2}). After removing the cache line from its own cache (\circled{3}), $N_2$ informs the directory (\circled{4}), which then updates the cache line's permissions to \textbf{M}, records $N_1$ as the owner (\circled{5}), and acknowledges $N_1$ with the cache line data (\circled{6}). $N_1$ then accesses the cache line (\circled{7}). 

While the above example demonstrates the \textbf{S}$\rightarrow$\textbf{M} transition in cache permission, other transitions are either similar or simpler. Specifically, \textbf{M}$\rightarrow$\textbf{M} and \textbf{M}$\rightarrow$\textbf{S} transitions require similar invalidations for the node initially holding the cache line with \textbf{M} permission. On the other hand, \textbf{S}$\rightarrow$\textbf{S}, \textbf{I}$\rightarrow$\textbf{S} and \textbf{I}$\rightarrow$\textbf{M} transitions require no invalidations, only updates to the sharer list and permissions for the cache line at the directory, while the data can directly be fetched from memory.


\paragraphb{Lock-based synchronization} Locks are crucial to the scalability of multi-threaded applications on large-scale shared memory systems. At a high level, a lock permits either a single exclusive writer or multiple concurrent readers to access a critical section. On multi-core and NUMA machines, the pursuit of high-performance locks has yielded numerous scalable algorithms atop hardware-based cache coherent substrates. For example, queue-based lock algorithms such as MCS~\cite{mcslock, mcsrwlock} and CLH~\cite{clhlock} achieve scalability by letting each requestor thread spin on a core-private cache line. Reader-writer locks can partition reader-indicators~\cite{linuxbrlock} to enable scaling of concurrent readers. On NUMA architectures, memory-hierarchy-aware locks~\cite{cloflock, cohortlock, hmcslock, cnalock, shufflelock} further improve locking scalability by exploiting memory locality.

We use the MCS lock algorithm~\cite{mcslock} as a concrete example of scalable lock algorithms. Consider a number of threads each running on its own CPU core with its own private cache, trying to acquire the same lock. The MCS lock queues requestor threads, with each queue entry residing in a separate cache line with two pieces of information --- an atomic flag (\code{waiting}), which tracks whether the requestor is waiting on the lock or not, and a pointer to the next queue entry (\code{next}). The requestor at the head of the queue holds the lock (\ie, its \code{waiting} is \code{false}), while other requestors spin locally on their queue entry --- specifically, the \code{waiting} flag, which is set to \code{true}. To release the lock, the lock holder (at the head of the queue) simply sets the \code{waiting} flag for the next requestor in the queue to \code{false}. Since all inter-cache communications are restricted to adjacent requestors in the queue in a fixed order, the MCS lock's inter-cache communications per lock acquisition and release are limited to a fixed constant, independent of the number of requestors.

\begin{figure}[t]
\centering
    \includegraphics[width=0.4\textwidth]{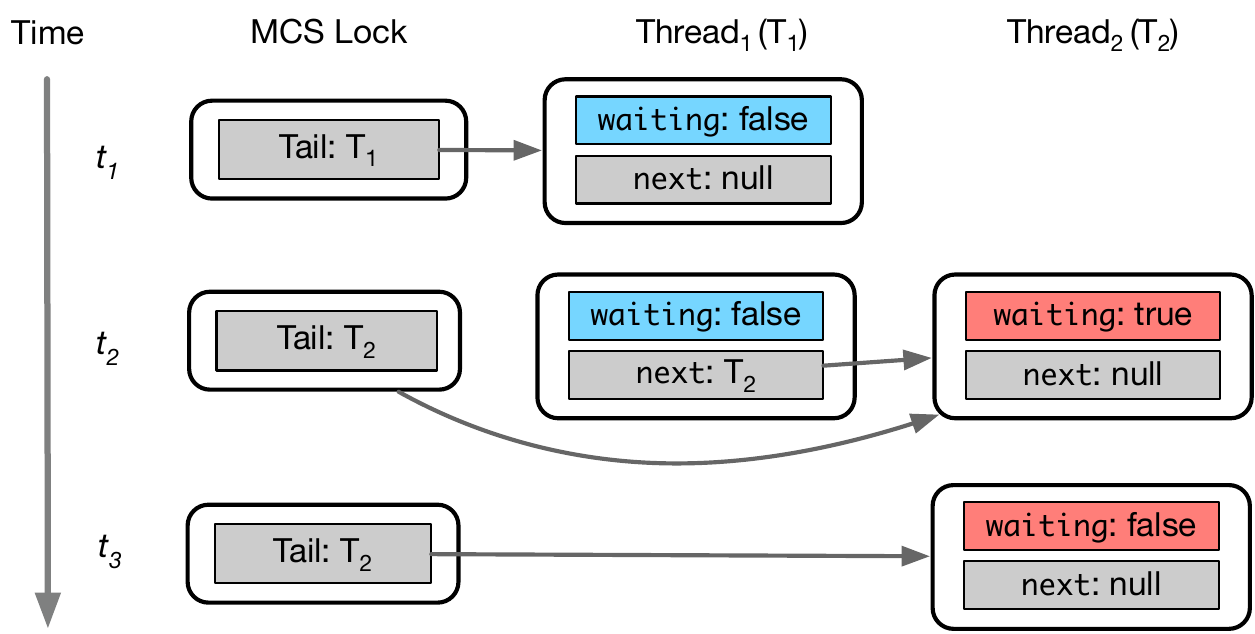}
  \vspace{-.5em}
  \caption{\textbf{MCS-Lock operation (\S\ref{ssec:background})} illustrated as the state of the queue at three times: $t_1$, $t_2$, and $t_3$.} 
  \label{fig:mcslock}
  \vspace{-1.5em}
%
\end{figure}


The example in Fig.~\ref{fig:mcslock} illustrates the operation of the MCS lock. At time $t_1$, thread $T_1$ acquires the lock as it is the only requestor in the queue. At $t_2$, $T_2$ is added to the queue, and polls at its private \code{waiting} flag until the value becomes \textit{false}. At $t_3$, $T_1$ hands over the lock to $T_2$ by setting $T_2$'s \code{waiting} as \code{false}. $T_2$ then detects its \code{waiting} is \code{false} and proceeds to its critical section.

\subsection{Inefficiencies due to a Layered Design}
\label{ssec:layereddesign}

Unfortunately, even well-optimized lock algorithms still trigger redundant cache-coherence transactions when layered atop cache-coherent substrates, resulting in additional lock handover latency and wasted interconnect bandwidth. We demonstrate this through an empirical evaluation of representative lock algorithms like MCS lock~\cite{mcslock}, \code{pthread} reader-writer lock~\cite{pthreadrwlock} and percpu reader-writer lock~\cite{linuxbrlock} in \S\ref{ssec:understandperf}. Ideally, each lock acquisition needs only one cache-coherence transaction (as explained in \S\ref{ssec:generalizedcc}). However, we find that even state-of-the-art lock algorithms trigger significantly more transactions and consequently observe millisecond-level lock acquisition latencies (Fig.~\ref{fig:eval_micro_perf}).

To better understand how the layered approach leads to such redundant communications, we once again use the workflow of the MCS lock algorithm as an example (Fig.~\ref{fig:mcslock}). In this example, the queue entries of threads $T_1$ and $T_2$ reside in two separate cache lines, $C_1$ and $C_2$, respectively, for scalability. Between time $t_2$ and $t_3$, $C_1$ and $C_2$ are both cached by $T_2$ with \textbf{M} permission because $T_2$ modified $T_1$'s {\code{next}} and created its own queue entry at $t_2$. When the lock ownership is transferred from $T_1$ to $T_2$ at $t_3$, it triggers $3$ sequential cache-coherence transactions in the MSI protocol:

\begin{enumerate}[itemsep=0pt, leftmargin=*]
  \item To find the next requestor,  $T_1$ fetches $C_1$, which contains the \code{next} field within its queue entry with \textbf{S} permission since it has previously been cached by $T_2$ with \textbf{M} permission at time $t_2$ --- this \textbf{M}$\rightarrow$\textbf{S} transition triggers an invalidation, as discussed in \S\ref{ssec:background};
  
  \item When $T_1$ updates $T_2$'s \code{waiting} to \code{false}, it must fetch $C_2$ with \textbf{M} permission, since it had previously been cached by $T_2$ with \textbf{M} permission at $t_2$ --- again, this \textbf{M}$\rightarrow$\textbf{M} transition triggers an invalidation;\label{step:2}
  
  \item Finally, $T_2$ can only detect that it owns the lock now after it reads its own queue entry's \code{waiting} flag. This requires fetching $C_2$ back with \textbf{S} permission since it was just cached by $T_1$ in step~\ref{step:2}. This \textbf{M}$\rightarrow$\textbf{S} transition triggers yet another invalidation.
\end{enumerate}
\noindent
In addition to the $3$ coherence transactions above, adding $T_2$ to the queue triggers $2$ more coherence messages. While the delays caused by these messages can be subsumed by the time that $T_2$ waits for the lock, the traffic still contributes to inefficient use of the inter-cache interconnect bandwidth. As such, the MCS lock incurs $5$ coherence transactions for each lock handover, with $3$ of them in the critical path.

The inefficiency is not limited to the MCS lock but is fundamental to all scalable lock algorithms. Specifically, scalable lock algorithms must partition the lock's state across multiple cache lines to restrict inter-cache communications. In the above example, the queue entries of $T_1$ and $T_2$ reside in two separate cache lines, $C_1$ and $C_2$, respectively. This unavoidably causes multiple sequential accesses to the cache lines containing the partitioned lock state for every synchronization operation, each of which triggers a distinct cache-coherence transaction. We note that these inefficiencies essentially stem from the layering of lock algorithms atop the shared memory abstraction, which uses cache coherence as a black box. A synchronization mechanism that is co-designed with the cache-coherence protocol can (\S\ref{sec:design}) --- and, as we demonstrate in \S\ref{sec:evaluation}, does --- circumvent these inefficiencies. In particular, we show that the acquisition and release of reader-writer locks can be facilitated with a \textit{single} coherence transaction.

It is understandable why such a co-design has not been explored for multi-core and NUMA architectures in the past --- given the low-latency, high-bandwidth cache-coherence substrates in multi-core and NUMA architectures, the additional inefficiency is all but negligible. As such, much of prior work has focused on improving the scalability of such locks, without considering the latency and number of interconnect messages triggered during their acquisition and release. However, with each inter-cache communication incurring several microseconds of delay and consuming a non-trivial fraction of the interconnect bandwidth in disaggregated architectures, this inefficiency often results in significant application performance degradation. 

\subsection{Lock services in Software DSMs}
\label{ssec:softwaredsm}

To circumvent the inefficiencies of layering synchronization atop cache-coherence substrates, software-based DSM systems like TreadMarks~\cite{1994kelehertWTC-treadmarks}, Munin~\cite{munin}, Midway~\cite{midway} and GAM~\cite{gam} employ weak memory consistency models~\cite{1992keleherISCA-releaseconsistency}) along with standalone software-based lock services that \textit{bypass} the cache-coherence layer altogether. 
For example, TreadMarks runs software-based lock managers to serve lock requests under a client/server model, while Munin~\cite{munin} and Midway~\cite{midway} implement lock objects with software-based distributed queueing protocols that bypass shared memory. In contrast, state-of-the-art disaggregated shared memory systems~\cite{mind, cxl} employ hardware cache-coherence interconnects for performance. While a software-based realization of lock services precludes microsecond-level performance, porting them to hardware interconnects would incur significant additional complexity and resources, which tend to be quite limited~\cite{mind}. As such, we focus on instead exploring if insights from these specialized lock services can be directly realized within the cache-coherence substrate with minimal modifications for minimal additional complexity and resource usage, as well as microsecond-level performance.

\section{\designname Design}
\label{sec:design}

We now describe \designname, a generalized cache-coherence protocol that provides lock-based synchronization semantics directly at the cache-coherence layer. \designname builds on our observation that synchronization is, in fact, a \emph{generalization} of cache coherence. We describe the required extensions to the cache-coherence protocol for achieving such a generalization in \S\ref{ssec:generalizedcc} and demonstrate how standard synchronization interfaces can be realized via our generalized cache-coherence protocol in \S\ref{ssec:designapi}. We conclude by describing additional optimizations enabled by our generalization for lock-based synchronization in \S\ref{ssec:designoptimizations}.

\subsection{Generalized Cache-coherence}
\label{ssec:generalizedcc}

Fundamentally, both cache coherence and lock-based synchronization strive for the same goal --- single-writer-multi-reader (SWMR) invariant~\cite{nagarajan2020primer} over some shared state. Under SWMR invariant, at any point in time, either a single entity that intends to modify the shared state has \emph{exclusive} access to it or multiple entities that intend to only read from the state without modifying it, have \emph{shared} access to it. SWMR invariant is the building block for ensuring correctness in layers above, \eg, cache state for cache coherence, and data consistency in synchronization. 

The key distinction between cache coherence and synchronization, however, stems from the temporal and spatial granularity at which SWMR invariant is enforced. In particular, cache-coherence protocols enforce it at a single instruction granularity (in time) and at a fixed cache line granularity (in space). In contrast, lock-based synchronization strives for SWMR invariant at arbitrary instruction count (referred to as a critical section) and arbitrary data size granularities. It is easy to see, then, that lock-based synchronization is simply a generalization of cache coherence in time and space. Historically, cache coherence has been implemented in hardware across multi-core caches and is invisible to software, where synchronization is implemented. As such, the latter is forced to recreate its generalized SWMR invariant atop the former.

With extensions to cache-coherence substrates, however, we can realize generalized cache-coherence protocols that can natively support lock primitives, making each lock acquisition a single cache-coherence transaction.
Our key design principle in this generalization is to identify the minimal set of extensions to existing cache-coherence protocols for two main reasons. First, this permits a resource-efficient realization with minimal additional complexity in programmable cache coherent substrates, which tend to be quite resource-constrained~\cite{mind}. Second, the smaller set of extensions makes for easier adoption in future hardware interconnects like CXL~\cite{cxl} or even multi-core architectures with a large number of cores (\S\ref{sec:futurework}).
We describe these minimal extensions next, namely the wait queue (\S\ref{ssec:designwaitqueue}) and the shared memory list (\S\ref{ssec:designvarsizecacheblock}), using the MSI cache-coherence protocol as our base protocol. Although their low-level implementation could be architecture-specific (\S\ref{sec:impl}), these extensions are applicable to any directory-based cache-coherence substrate that can support such extensions.
Note that while similar generalizations are possible for more complex protocols~\cite{mesi,mosi,moesi}), we focus on the MSI protocol for its simplicity.

\subsubsection{Wait queue for Temporal Generalization}
\label{ssec:designwaitqueue}

As we saw in \S\ref{ssec:background}, a directory-based cache-coherence protocol ensures SWMR invariant by tracking the permission of the requested cache line --- \textbf{M}, \textbf{S} or \textbf{I}. If an instruction's execution at one node (\eg, a CPU core in multi-core architecture or a compute blade in disaggregated architectures, \etc) requests a cache line with specific permission, the protocol triggers a transaction that makes the cache line available with that permission to that instruction immediately. For instance, in Fig.~\ref{fig:cache_coherence_example}, the protocol provides the cache line to node $N_1$ with $M$ permission via a transaction that (i) invalidates the cache line at $N_2$'s cache, and (ii) updates the sharer list and permission for the cache line at the directory.

\begin{figure}[t!]
\centering
\includegraphics[width=0.35\textwidth]{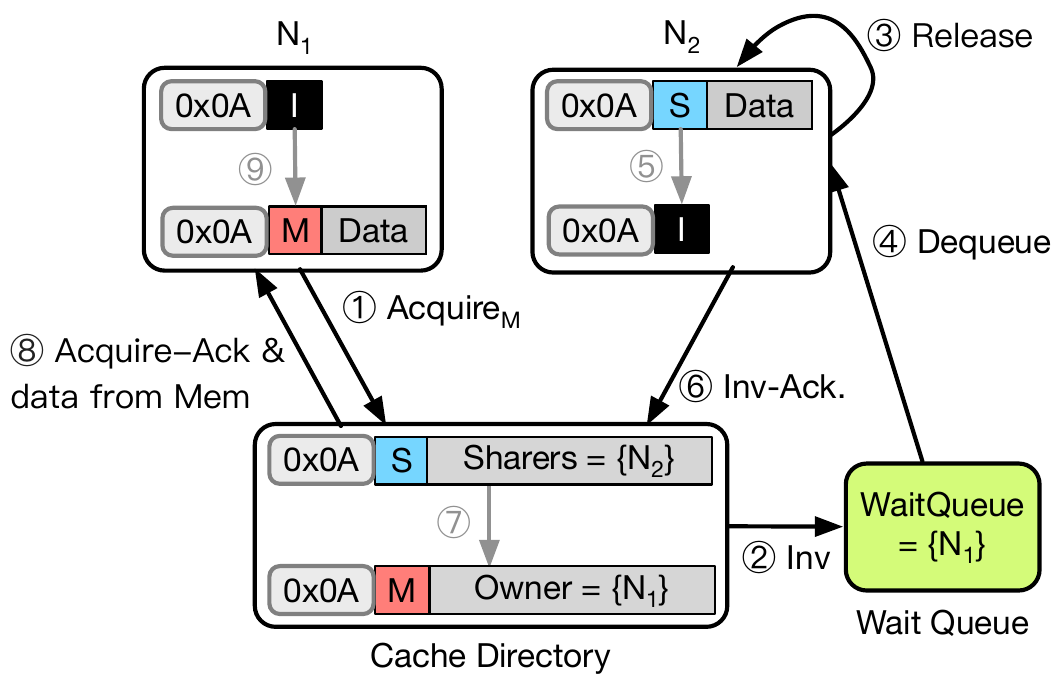}
\vspace{-10pt}
\caption{\textbf{Temporal generalization with wait queues~(\S\ref{ssec:designwaitqueue}).}}
\label{fig:gcs_wait_queue_example}\vspace{-1.5em}
\end{figure}

To enable the temporal generalization of the protocol, where a node can hold the cache line with certain permission for an arbitrary number of instructions (critical section), we add two new request types, \code{Acquire}~(\textcircled{1} in Fig.~\ref{fig:gcs_wait_queue_example}) to request the cache line mark the beginning of the critical section and \code{Release}~(\textcircled{3} in Fig.~\ref{fig:gcs_wait_queue_example}) to release the cache line and explicitly mark the end of the critical section. Moreover, we must be able to \emph{delay} other nodes from acquiring the cache line immediately in some cases. Specifically, other nodes should be able to acquire the cache line only after the first node explicitly releases it. A natural way to enable such deferred cache line (and associated permission) transfers is by enqueuing requests to the cache line in a \emph{wait queue}, and dequeuing a request only when the first node releases the cache line. This is akin to wait queues used in synchronization primitives, except the queue is embedded within the cache-coherence layer.

Fig.~ \ref{fig:gcs_wait_queue_example} demonstrates how a wait queue can enable temporal generalization for the same example in Fig~\ref{fig:cache_coherence_example}. The target cache line is initially cached at $N_2$ with \textbf{S} permission. $N_1$ then issues an \code{Acquire} request for the same cache line with \textbf{M} permission to the cache directory (\circled{1}). The directory looks up the current cache line permission (\textbf{S}) and sharer list (\{\textbf{$N_1$}\}), realizing that the request requires $N_2$ to relinquish the cache line via invalidation. In contrast to standard MSI protocol execution, the directory defers the invalidation and instead enqueues $N_1$'s request in a wait queue associated with the cache line (\circled{2}). Only when $N_2$ finishes its critical section and voluntarily releases the cache line via \code{Release} (\circled{3}) is the request dequeued (\circled{4}) and the invalidation performed at $N_2$ (\circled{5}). The remainder of the cache-coherence transaction proceeds as per the standard MSI protocol --- $N_2$ informs the directory (\circled{6}), which updates the cache line's permission to \textbf{M} and marks $N_1$ as the only sharer (\circled{7}), and sends $N_1$ an acknowledgment along with the cache line data (\circled{8}).  

As with cache coherence, other permission transitions are either similar or simpler. Specifically, \textbf{M}$\rightarrow$\textbf{M} and \textbf{M}$\rightarrow$\textbf{S} transfers require similar deferred invalidations by enqueuing the transfer requests at the wait queue until the node holding the cache line with \textbf{M} permission explicitly releases it. Moreover, \textbf{S}$\rightarrow$\textbf{S} transfers do not require enqueuing the request since multiple readers can hold the cache line simultaneously under the SWMR invariant. Similarly, \textbf{I}$\rightarrow$\textbf{S} and \textbf{I}$\rightarrow$\textbf{M} transfers also do not require enqueueing requests, since no node has the cache line to begin with. 


Note that the location of the wait queue does not affect correctness, but does affect performance. We defer the discussion on how our implementation navigates various tradeoffs for wait queue placement to \S\ref{ssec:implwaitqueue}.

\subsubsection{Shared memory list for Spatial Generalization}
\label{ssec:designvarsizecacheblock}
Unlike coherence protocols that track the permission for a fixed-size cache line, synchronization must preserve SWMR invariant for arbitrary amounts of shared state. The shared state may be fragmented, and may even be empty. This requires an extension to cache-coherence protocols to track multiple shared memory locations of arbitrary size. Specifically, instead of being a single address with a fixed size, each address tag in \designname is a list of $(m_i, s_i)$ pairs, where $m_i$ and $s_i$ are the base address and size in bytes of a shared memory region, respectively. For standard cache coherence, this list simply reduces to a single entry of $64$ B.

\begin{figure}[t!]
\centering
\vspace{-5pt}
\includegraphics[width=0.35\textwidth]{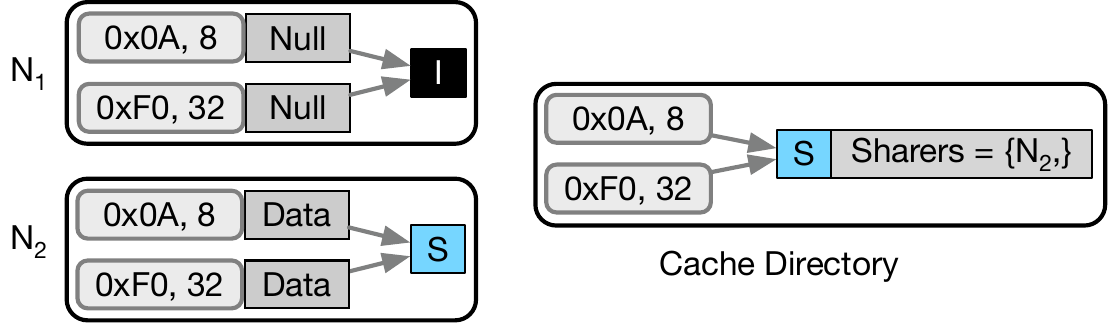}
\vspace{-10pt}
\caption{\textbf{Spatial generalization with shared memory lists~(\S\ref{ssec:designvarsizecacheblock}).} 
}
\label{fig:gcs_shmlist_example}\vspace{-1.5em}
\end{figure}

Fig.~\ref{fig:gcs_shmlist_example} shows our spatial generalization for the same example as Fig.~\ref{fig:gcs_wait_queue_example}. The directory and line together track two shared memory regions: \{$(\code{0x0A}, 8), (\code{0xF0}, 32)$\}. The protocol execution is identical to the description in \S\ref{ssec:designwaitqueue}, except the invalidation step \circled{5} now requires $N_2$ to remove two memory regions of different sizes from its cache, and in step \circled{8}, the directory sends $N_1$ the data corresponding to both regions along with the acknowledgment. Again, we defer a discussion of the tradeoffs stemming from an architecture-specific implementation of the shared memory list to \S\ref{ssec:implmultilocationentry}.

\subsubsection{Impact on Memory Consistency}
Since all of our modifications are confined to the cache-coherence protocol, \designname does not impose any restrictions on the memory consistency model, \ie, it can work with any consistency model ranging from more relaxed release consistency models to stricter TSO consistency models. For instance, our implementation of \designname atop disaggregated memory works with a TSO consistency model.

%
%

\begin{table}[t]
    \centering
    \tiny
    \resizebox{0.48\textwidth}{!}{
    \begin{tabular}{l|l}
      \hline
      \textbf{Language}  & \textbf{Code Snippets}\\\hline\hline
      C (\scode{pthread})  & \specialcell{
        \scode{pthread\_rwlock\_t l;}\\
        \scode{pthread\_rwlock\_init(\&l, NULL); \textcolor{magenta}{// Initialize lock}}\\
        \hline
        \scode{pthread\_rwlock\_wrlock(\&l);}\\
        \scode{\textcolor{magenta}{// Critical section for write lock}}\\
        \scode{pthread\_rwlock\_unlock(\&l);}\\
        \hline
        \scode{pthread\_rwlock\_rdlock(\&l);}\\
        \scode{\textcolor{magenta}{// Critical section for read lock}}\\
        \scode{pthread\_rwlock\_unlock(\&l);}
        }\\\hline
      Rust & \specialcell{
        \scode{struct Data \{}\\
        \quad\scode{key: u64,}\\
        \quad\scode{val: u64}\\
        \scode{\}}\\
        \scode{\textcolor{magenta}{// Initialize lock}}\\
        \scode{let l = RwLock<Data>::new(Data::default());}\\
        \hline
        \scode{\{ \textcolor{magenta}{// Critical section for write lock}}\\
        \quad\scode{let mut w = l.write().unwrap();}\\
        \quad\scode{(*w).key = 42;}\\
        \quad\scode{(*w).val = 42;}\\
        \scode{\}}\\
        \hline
        \scode{\{ \textcolor{magenta}{// Critical section for read lock}}\\
        \quad\scode{let r = l.read().unwrap();}\\
        \quad\scode{assert\_eq!((*r).key, 42)}\\
        \quad\scode{assert\_eq!((*r).val, 42)}\\
        \scode{\}}
      } \\
     \hline
    \end{tabular}
    }
    \vspace{-5pt}
    \caption{\textbf{Lock-based synchronization in C and Rust~(\S\ref{ssec:designapi}).}}
    \label{tab:designapi}
    \vspace{-4em}
\end{table}

\subsection{Supporting Synchronization Interfaces}
\label{ssec:designapi}

While \S\ref{ssec:generalizedcc} described how the directory-based MSI cache-coherence protocol can be generalized, we now describe how various popular synchronization interfaces can leverage the generalized protocol for efficient performance scaling. For completeness, we discuss the adaptation of \designname to two popular lock-based synchronization interfaces with slightly different expressiveness.

\paragraphb{Synchronization in C (\code{pthread})} Arguably, the most popular interface for lock-based synchronization in multi-threaded applications is the \code{rwlock} in the POSIX threads (\code{pthread}) library interface (Table~\ref{tab:designapi}). The lock is acquired in either write or read mode via \code{pthread\_rwlock\_wrlock} or \code{pthread\_rwlock\_rdlock}, respectively, and released via \code{pthread\_rwlock\_unlock}. In generalized cache coherence, the cache line simply tracks the address of the lock variable (\code{l}). Acquiring the lock in write or read mode triggers a request for the cache line with \textbf{M} or \textbf{S} permission, respectively. Finally, releasing the lock triggers releasing the cache line in the generalized cache-coherence protocol. Note that the POSIX API does not explicitly specify the shared memory that will be accessed in the critical section, excluding the opportunity of combining data fetching with lock acquisition for improved performance. As such, we also support more expressive APIs as described next.

\paragraphb{Synchronization in Rust (\code{std::sync})} Table~\ref{tab:designapi} also shows the operation of Rust's \code{std::sync::RwLock}. Unlike \code{pthread}'s lock-based synchronization, \code{RwLock} explicitly takes the state being protected by the lock (a \code{Data} object in our example) as an argument. This makes adapting \designname to \code{RwLock} even simpler and more efficient. In particular,
the cache line and directory entry can simply track the shared memory list containing the address and size of the shared object (\eg, the protected \code{Data} object in the shown example).
Lock acquisition and release, on the other hand, proceed similarly to \code{pthread}, as described above.

\paragraphb{A note on cache-lines not used as locks} We note that while both the lock realizations above simply leverage some cache lines as locks, the temporal and spatial generalizations introduced in \designname do not affect the operation of cache coherence for regular cache entries --- they still operate at a single cache line and single instruction granularity (\ie, they do not use the wait queue introduced in \S\ref{ssec:designwaitqueue}).

\subsection{Cache-coherence Optimizations}
\label{ssec:designoptimizations}

%


As one would expect, our generalized cache-coherence protocol achieves performance scalability because it avoids redundant communications observed in a layered design (\S\ref{ssec:background}). Indeed, our analysis of the MCS lock workflow in \S\ref{ssec:layereddesign} showed that the critical path in lock-based synchronization --- the lock handover to the next requestor --- requires three sequential cache-coherence transactions. In contrast, our generalized cache-coherence protocol provides the same SWMR invariant as the MCS lock but with the lock handover requiring a single cache-coherence transaction (\S\ref{ssec:designwaitqueue}).

Interestingly, our generalized cache coherence also has the fortunate side-effect of inheriting optimizations from traditional cache-coherence protocols. These optimizations can further reduce overheads in many common scenarios seen in lock-based synchronization, as we demonstrate below.

\paragraphb{Acquiring shared state along with lock} As shown in \S\ref{ssec:designapi}, popular lock-based synchronization interfaces (\eg, \code{pthread}) decouple lock acquisition from fetching the shared data associated with the lock, which results in additional delay and coherence traffic in placing the corresponding shared state in the requestor's cache. Although modern languages like Rust address this issue to some extent by coupling the shared state with the lock, the layering of lock atop cache coherence provides no guarantee that the underlying hardware will fetch both together. Moreover, since cache line sizes are limited to $64$ B in traditional architectures, placing any fragmented shared state, or state of size greater than $64$ B in the requestor's cache requires multiple cache-coherence transactions. In contrast, our generalized protocol performs the acquisition of the shared state (cache line) and the lock (access permission) in a single transaction, akin to traditional cache coherence. Moreover, with our spatial generalization, we can place all shared states protected by a lock (of any size and with any amount of fragmentation) within the requestor's cache, in a single cache-coherence transaction.



\paragraphb{Exploiting temporal locality for locks} In traditional cache-coherence protocols, once a cache line is placed in a requestor's cache, it remains there until it is invalidated, in order to exploit the temporal locality of data accesses. In extending such protocols, our generalized cache coherence inherits the same optimization --- both the lock and the shared state associated with it remain in the requestor's cache until it is explicitly invalidated by another request. Interestingly, optimized lock algorithms in multi-core architectures do exploit a similar optimization, wherein threads running on the same core can enter the critical section multiple times without communicating with other cores, as long as threads on other cores do not attempt to acquire the same lock. However, standalone lock services~\cite{lockservice1, lockservice2, lockservice3, netlock} cannot exploit such an optimization because their locks are decoupled from shared memory.

\section{\name: A \designname Implementation}
\label{sec:impl}
\vspace{-0.25em}

We present \name, an implementation of \designname for disaggregated shared memory that supports both \designname APIs~(detailed in \S\ref{ssec:designapi}) for efficient lock-based synchronization. We begin with a
brief background on the shared disaggregated memory platform that \name builds on (\S\ref{ssec:platform}). We then describe design details on how we incorporate wait queues using a novel queue transfer protocol (\S\ref{ssec:implwaitqueue}) and shared memory lists (\S\ref{ssec:implmultilocationentry}) into MIND's cache-coherence adhering to the constraints imposed by programmable hardware.
%


\subsection{Disaggregated Shared Memory Platform}
\label{ssec:platform}

While generalized cache coherence can be realized on any flexible cache-coherence substrate, \name focuses on disaggregated architectures, since their higher latency and lower bandwidths present a pressing need for more efficient and scalable synchronization primitives (\S\ref{ssec:layereddesign}). 

A recent approach for rack-scale disaggregated shared memory, MIND~\cite{mind}, realizes an efficient directory-based MSI cache-coherence substrate by leveraging in-network processing. We use MIND as the underlying framework for \name for three reasons. First, since much of the cache-coherence logic is implemented in hardware, it observes better scalability and lower latency relative to software-based alternatives. Second, since MIND is implemented across P4 programmable network hardware and the Linux kernel, it is extensible enough to support our cache-coherence generalizations. Finally, MIND is publicly available~\cite{mind-github}. We discuss other potential platforms for \designname implementation in \S\ref{sec:futurework}.

\begin{figure}[!t]
  \centering
  \includegraphics[width=0.45\textwidth]{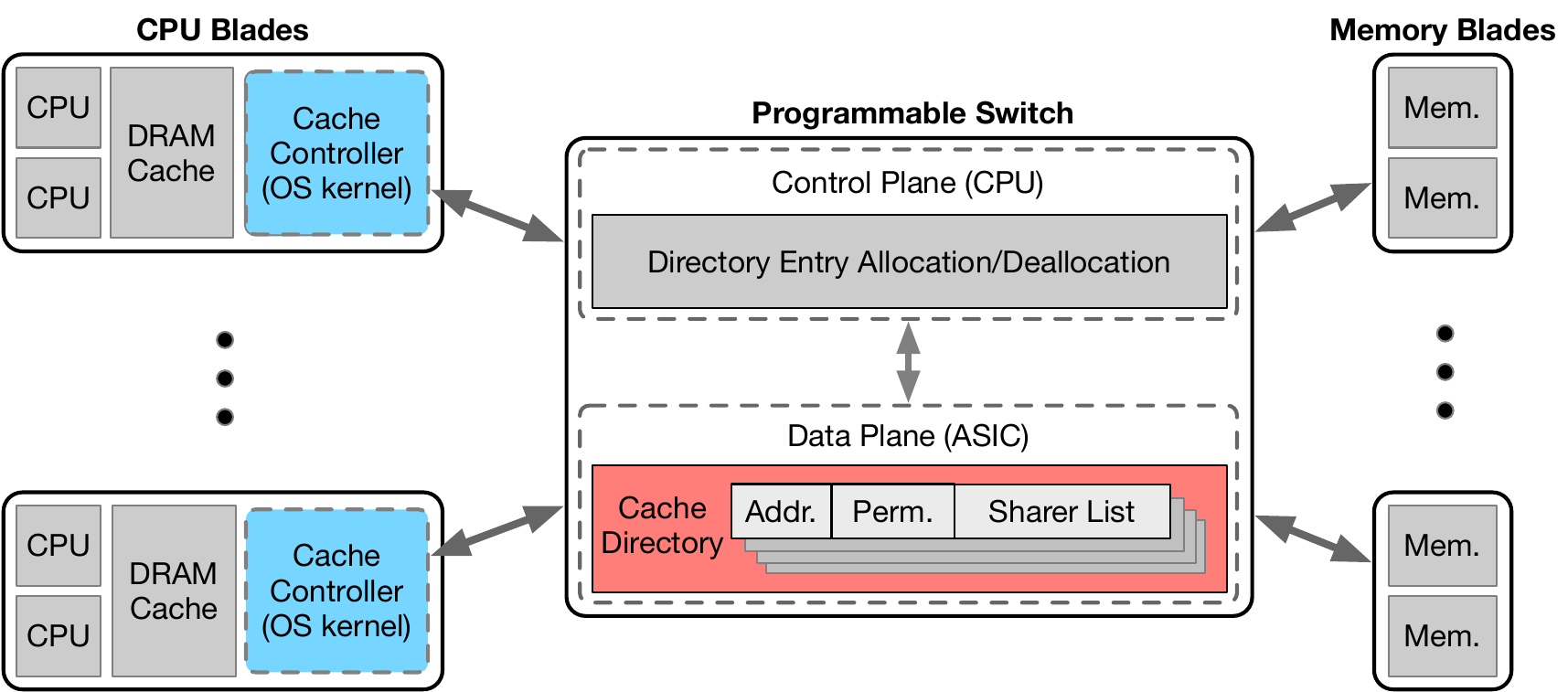}
  \vspace{-5pt}
  \caption{\textbf{In-network Directory-based Cache Coherence Architecture in MIND (\S\ref{ssec:platform}).} \name's modifications are confined to the cache controller (blue) and the cache directory (red).}
  \label{fig:implmind}
  \vspace{-1.5em}
\end{figure}

\paragraphb{MIND Architecture} Fig.~\ref{fig:implmind} shows MIND's rack-scale architecture for cache coherence. It comprises compute and memory blades connected via a programmable network switch. Each compute blade is equipped with a small amount of DRAM used as a cache --- if an application accesses a cache line (page granularity in MIND) not present in the DRAM cache, it triggers a page fault. The fault handler initiates coherence transactions via cache controller logic implemented in the kernel --- it sends out requests for the cache line with \textbf{S} or \textbf{M} permissions for faulting \code{LOAD} or \code{STORE} operations, respectively. The programmable switch implements the cache directory, issuing invalidations to relevant compute blades and updating its local cache line state (permissions and the sharer list) as necessary. Invalidation requests are handled by the in-kernel cache controller logic at compute blades, which removes the cache lines from the DRAM cache, triggering both CPU cache and TLB invalidations to guarantee consistency between the CPU's private caches and the DRAM cache. In case none of the compute blades have the cache line in their cache, the data is fetched from disaggregated memory via RDMA. While MIND implements additional components for realizing a complete virtual memory subsystem, we omit their details since they are unnecessary for understanding \name.


While MIND's realization of cache coherence across programmable network hardware and kernel software affords both performance and flexibility, the in-network implementation of its cache directory also imposes resource constraints on extensions to the directory. Specifically, our generalizations to cache coherence (\S\ref{ssec:generalizedcc}) require additional storage and processing logic for every cache line, ideally at the directory. However, the programmable switch ASIC only has a few megabytes of on-chip memory and can only support a few cycles of computation per packet~\cite{mind}, much of which is already used up by MIND. Therefore, \name's implementation atop MIND must navigate various tradeoffs between feasibility and efficiency for realizing the wait queues and shared memory lists, as we discuss next. 

\subsection{Implementing the Wait Queue}
\label{ssec:implwaitqueue}

We begin by highlighting the challenges of implementing the wait queue and then describe our novel queue transfer protocol to address them. 

\subsubsection{Challenges} At first glance, the cache directory appears to be a good location for the wait queue --- its central location simplifies consistency issues for concurrent updates and makes it accessible from any compute blade in half a round-trip. However, the queue requires a non-trivial amount of already scant storage and processing resources at the switch for enqueue and dequeue operations. On the other hand, while a realization at the memory blades suffers none of the resource constraints, it incurs additional network delays to communicate with the memory blade for each cache-coherence transaction --- defeating our main goal of eliminating unnecessary network traffic for better performance.


Placing the wait queue at the compute blades circumvents both the resource limitations of a switch-based realization and the performance overheads of a shared memory realization. However, one challenge that remains is to minimize network delays to facilitate the change in ownership of the cache line, which might require moving the queue itself. In particular, when a compute blade releases a cache line, the next request for the cache line in the queue should ideally be processed without any network delays. A straightforward solution is to ensure the wait queue is present at the cache line's current owner(s) (\ie, either the single exclusive writer or multiple concurrent readers). However, this leads to another challenge: if multiple compute blades (readers) hold the cache line, the wait queue must be replicated across all readers so that they all can locally dequeue the next request, requiring a mechanism to ensure consistency across the replicated queues. The problem is further exacerbated when the set of compute blades (readers) that hold the cache line change over time.

\subsubsection{A novel queue transfer protocol} We address the above challenges with a novel queue transfer protocol between compute blades, which guarantees (i) no network delay in processing the next request in the queue, and (ii) consistent accesses to the queue itself. Moreover, it minimizes the amount of state and logic at the switch, since the cache directory only needs to track which compute blade holds the queue for each cache line (referred to as the queue holder), in order to forward corresponding access requests.

\begin{figure}[t!]
  \includegraphics[width=0.7\columnwidth]{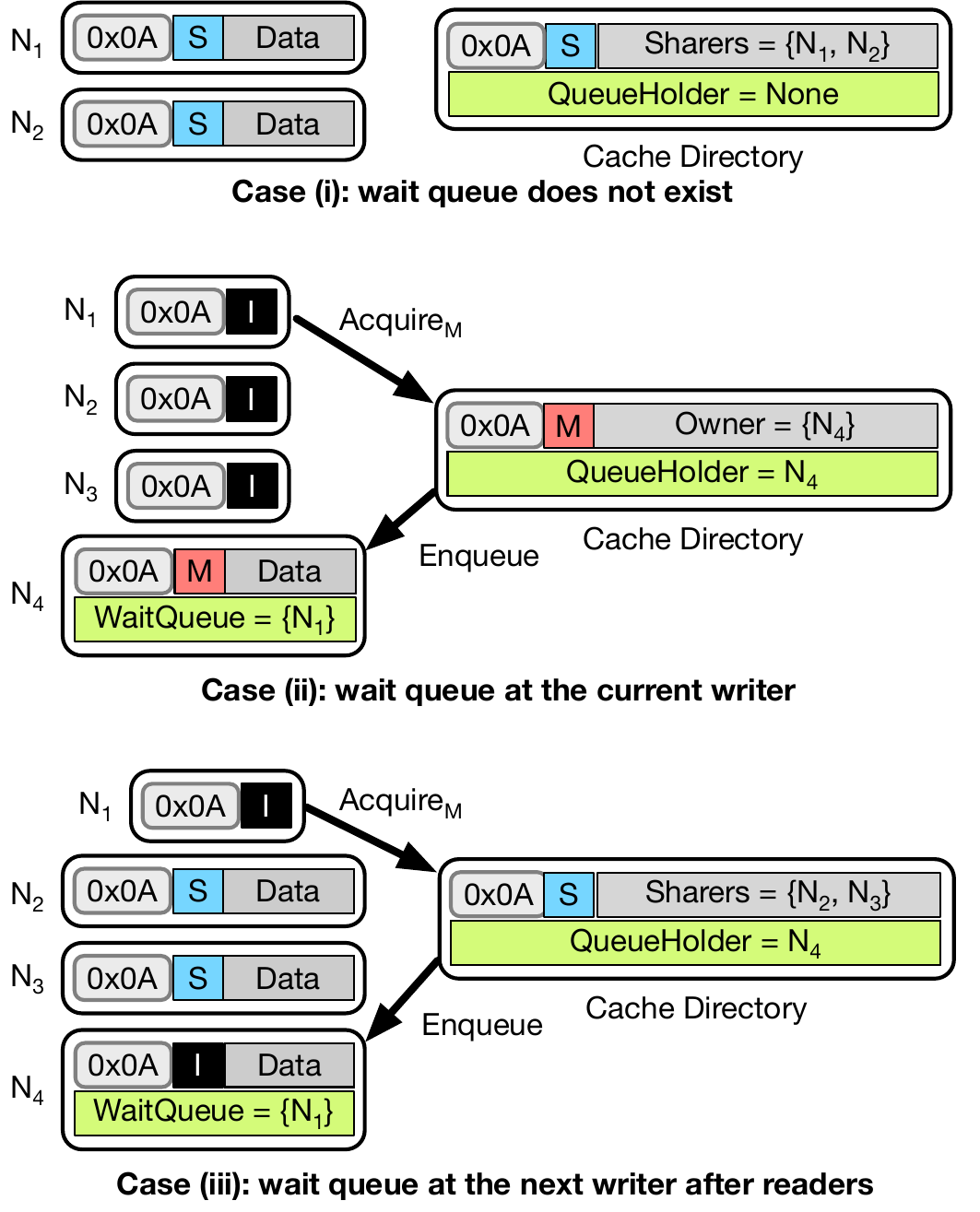}
  \vspace{-5pt}
  \caption{\textbf{Wait queue holders under different cases in \name ~(\S\ref{para:queue_holders}).} Case~(i): the wait queue does not exist without writers; Case~(ii): the wait queue is at the current writer; Case~(iii): the wait queue is at the next writer after current readers.
  }
  \label{fig:implqueueholder}\vspace{-1em}
\end{figure}

\paragraphb{Queue holders}\label{para:queue_holders} A queue holder enqueues any requests for a cache line until it is voluntarily released. \name ensures that there is only a single queue holder for a cache line at any point in time, avoiding any consistency issues associated with replicated queues. To achieve this, we leverage \designname's SWMR invariant: since only a single writer (\ie, a thread requesting the cache line with \textbf{M} permission) can hold a cache line at a given time, only one blade (the one hosting the writer) needs to track the queue at that time. While multiple readers (\ie, threads requesting the cache line with \textbf{S} permission) can hold a cache line concurrently, placing the cache line at additional readers does not require enqueueing requests (\S\ref{ssec:designwaitqueue}), \ie, no queue needs to be tracked for a cache line that is only requested by readers. However, if a cache line initially held by multiple readers is subsequently requested by a writer, a queue is created for it at the compute blade hosting the writer.

\noindent
Fig.~\ref{fig:implqueueholder} shows three possible cases for a cache line's wait queue: 
\begin{itemize}[leftmargin=*, itemsep=0pt]
\item{\textbf{Case (i)}} The wait queue does not exist when there are no writers requesting the cache line (\ie, for a cache line with \textbf{I} or \textbf{S} permission that has no waiting writers).
\item{\textbf{Case (ii)}} The wait queue is at the current writer (\ie, for a cache line with \textbf{M} permission) 
\item{\textbf{Case (iii)}} The wait queue is at the next writer when the cache line is held by one or more readers (\ie, for a cache line with \textbf{S} permission with a waiting writer).
\end{itemize}
\noindent
Additionally, the directory always maintains the sharer list and forwards invalidations as needed for \designname (\S\ref{ssec:designwaitqueue}). 

\paragraphb{Queue transfers} Algorithm~\ref{alg:impl_code_queue_transfer_protocol} describes the wait queue transfer protocol. When a writer attempts to acquire a cache line (\textcircled{1} in Fig.~\ref{fig:gcs_wait_queue_example}), either the wait queue has already transferred to it (described next) or it must create an empty one (transition to \textbf{Case (ii)}). When a writer releases the line (\textcircled{3} in Fig.~\ref{fig:gcs_wait_queue_example}), the writer drops the queue if there is no waiting requestor (line~\ref{line:algo1_drop}, transition to \textbf{Case (i)}). Otherwise, the writer processes the next queue entry and transfers the queue to the next writer (lines~\ref{line:algo1_non_empty_writer_start}--\ref{line:algo1_non_empty_writer_end}). Specifically, if the next requestor is a writer, the queue is transferred to it (transition to \textbf{Case (ii)}). If the next requestor is instead a reader (or multiple readers) and there is a writer waiting behind it (them), the queue is transferred directly to the writer (line~\ref{line:algo1_writer_behind_reader}, transition to \textbf{Case (iii)}). If there are no writers behind the reader(s), the queue is dropped (transition to \textbf{Case (i)}).
%

Note that since readers never hold the queue, the dequeue operation happens only in \textbf{Case (ii)}, when the current writer is the queue holder. This property ensures no network delays are incurred for dequeuing and processing the next requestor. A corner case that \name must handle is when multiple readers pass on the lock to a waiting writer; since the queue is already transferred to the writer in line~\ref{line:algo1_writer_behind_reader} of Algorithm~\ref{alg:impl_code_queue_transfer_protocol}, the readers must know who to transfer the cache line and its ownership to after they release it. To facilitate this, all readers are notified about the next waiting writer, and they notify the waiting writer (with an Inv-Ack message, \textcircled{6} in Fig.~\ref{fig:gcs_wait_queue_example}) whenever they release the cache line. Once the writer receives notifications from all readers in the sharer list, the directory entry is updated and the writer becomes the new owner. 




\newlength{\textfloatsepsave} \setlength{\textfloatsepsave}{\textfloatsep} \setlength{\textfloatsep}{0pt}
\begin{algorithm}[!t]
\caption{\textbf{Queue transfer protocol.}}
\label{alg:impl_code_queue_transfer_protocol}
\begin{algorithmic}[1]\footnotesize
\Procedure{On $Acquire$ at writer (\textcircled{1} in Fig.~\ref{fig:gcs_wait_queue_example})}{}
  \If{no $Queue$ was transferred to it}
    \State Initialize empty $Queue$ \Comment{to Case (ii)}\label{line:algo1_create}
  \EndIf
\EndProcedure
\Procedure{On $Release$ at writer (\textcircled{3} in Fig.~\ref{fig:gcs_wait_queue_example})}{}
\If{$Queue$ is empty}
  \State Drop the $Queue$ \Comment{to Case (i)}\label{line:algo1_drop}
\Else
  \State $nextRequestor \gets Queue.Dequeue()$\label{line:algo1_non_empty_writer_start}
  \If{$nextRequestor.permission$ == \textbf{M}}
    \State Transfer $Queue$ to $nextRequestor$ \Comment{to Case (ii)}\label{line:algo1_non_empty_writer_end}
  \ElsIf{$nextRequestor.permission$ == \textbf{S}}\label{line:algo1_non_empty_next_shared}
    \If{$Queue$ contains a writer}
      \State Transfer $Queue$ to next writer \Comment{to Case (iii)}\label{line:algo1_writer_behind_reader}
    \Else
      \State Drop the $Queue$ \Comment{to Case (i)}
    \EndIf
  \EndIf
\EndIf
\EndProcedure
\end{algorithmic}
\end{algorithm}
\setlength{\textfloatsep}{\textfloatsepsave}

\paragraphb{Consistency during queue transfers} The protocol must additionally consider a case where the cache directory forwards an access request to a queue holder that is in the process of transferring the queue to another compute blade: should such a request be processed by the original queue holder or the next one? To resolve such ambiguities, \name employs a versioning mechanism to ensure the queue transfer occurs \emph{atomically}, effectively ensuring that an access request is never forwarded to a compute blade in the middle of a queue transfer. Specifically, the directory maintains a version number for each cache line that tracks the number of access requests it has forwarded to the queue holder, while the queue holder maintains its own version number to track the number of access requests it has received from the directory. A wait queue transfer is approved by the switch only if the queue holder's version number matches that of the directory, ensuring that all access requests forwarded by the directory must have been processed at the queue holder before the holder initiated the transfer. If the switch denies a queue transfer, the queue holder reattempts the queue transfer after receiving the notification from the switch. On a successful transfer, the version numbers at both the switch and the queue holder are reset to zero.

\paragraphb{Bounded wait queue size} One concern regarding the wait queue transfers is the size of the queue: if the wait queue grows large under contention, it could incur significant performance overheads during transfers. However, \name's use of \designname is limited to inter-blade synchronization; as such, the length of the wait queue is bounded by the number of blades (tens of bytes in practice) rather than the number of threads in a cluster. We additionally implement a hierarchical locking layer for \name using the lock-cohorting technique~\cite{cohortlock, cloflock} to handle intra-blade contentions. 

\subsection{Implementing the Shared Memory List}
\label{ssec:implmultilocationentry}

To realize the spatial generalization under the limited memory and processing constraints of programmable switches, we decouple the shared memory list from the directory entries and maintain them at the compute blades, \ie, as metadata associated with a cache line in kernel software. Specifically, instead of tracking a single fixed-size region, each cache line in \name tracks multiple shared memory regions of arbitrary size with a list of $(m_i, s_i)$ pairs, where $m_i$ and $s_i$ are the base address and size in bytes of a shared memory region, respectively. On receiving invalidations, the shared memory list tracked by a line is invalidated atomically.

\section{Evaluation}
\label{sec:evaluation}


\begin{figure*}[ht!]
\centering
\vspace{-8pt}
\begin{minipage}{.57\textwidth}
  \includegraphics[width=\textwidth]{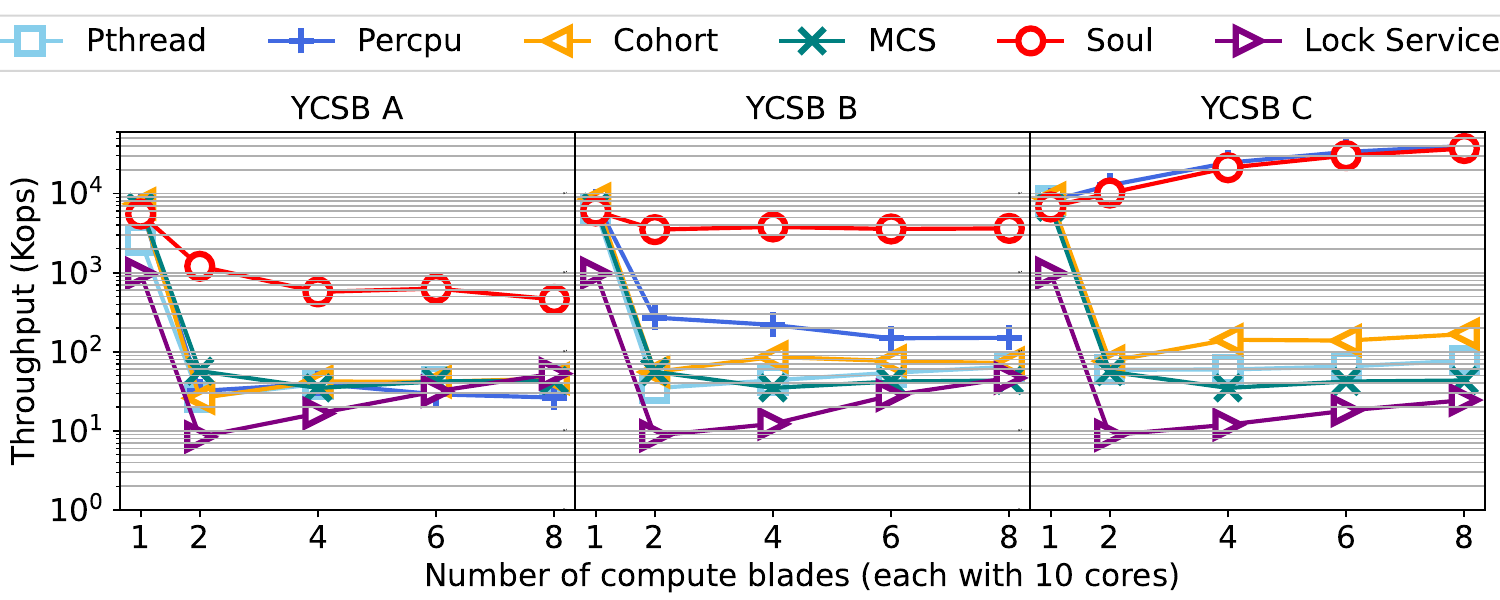}
  \captionof{figure}{\textbf{Performance scaling for MIND-KVS across YCSB A, B and C workloads~(\S\ref{ssec:eval_macro}).} \name outperforms the other compared systems enabling linear scaling, especially for YCSB-C, while the others suffer from inefficiencies due to layering atop cache coherence. Y-axis is in log scale.
  }
  \label{fig:eval_kvs}
\end{minipage}
\hfill
\begin{minipage}{.39\textwidth}
  \vspace{-1em}
  \includegraphics[width=\textwidth]{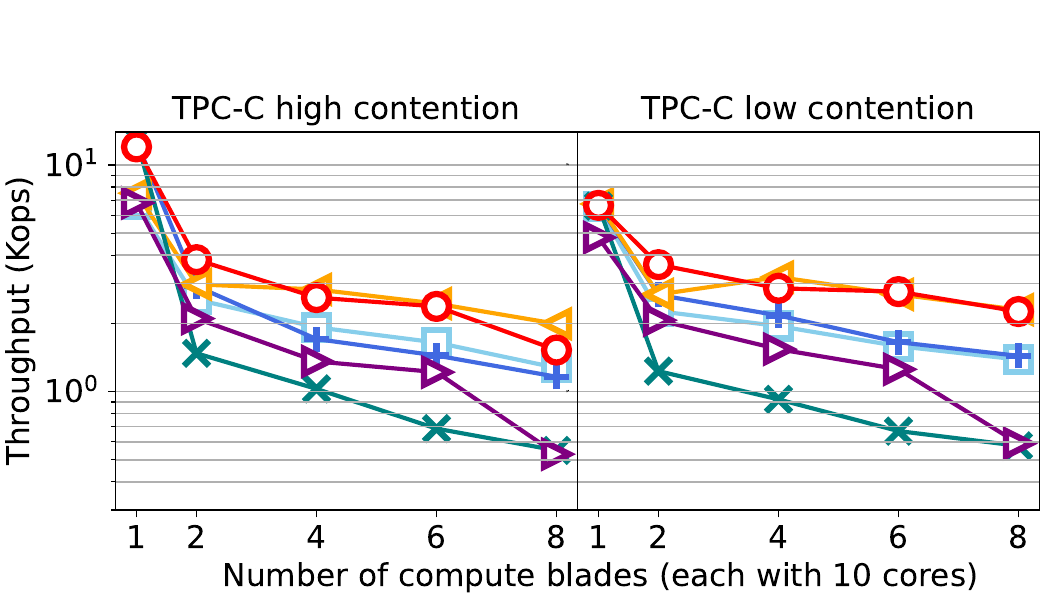}
  \captionof{figure}{\textbf{Performance scaling for Kyoto Cabinet with TPC-C workloads (\S\ref{ssec:eval_macro}).}
  Being memory-hierarchy aware, \name and Cohort observe the best performance; All systems observe performance degradation beyond 1 blade due to global lock design. Y-axis is in log scale.
  }
  \label{fig:eval_kc}
\end{minipage}
\vspace{-1em}
\end{figure*}




We evaluate \name's performance scaling for real-world applications on disaggregated memory (\S\ref{ssec:eval_macro}), its overheads (\S\ref{ssec:understandperf}), and the contributions of its optimizations (\S\ref{ssec:understandopt}).

\paragraphb{Compared systems} We compare \name against the two classes of approaches discussed in \S\ref{sec:motivation}. The first is state-of-the-art lock algorithms layered atop cache coherence; we consider multiple lock algorithms: (i) \texttt{MCS}~\cite{mcslock}, a representative of the queue-based mutual-exclusive (mutex) lock algorithms; (ii) \texttt{Pthread} (specifically, \texttt{pthread\_rwlock}~\cite{pthreadrwlock}), a reader-writer lock with centralized reader-indicators; (iii) \texttt{Percpu}, a reader-writer lock algorithm that implements fully decentralized (\ie~per-core) reader-indicators (similar to Linux's big reader lock~\cite{linuxbrlock}); (iv) \texttt{Cohort}, the \texttt{C-RW-WP} reader-writer lock~\cite{numarwlock}, a memory hierarchy-aware reader-writer lock. The second is a standalone lock service, \texttt{Lock-Service}, a software-based lock service that separates locking from the cache coherence for better efficiency~(\S\ref{ssec:softwaredsm}). We implemented the lock service following TreadMarks's design~\cite{1994kelehertWTC-treadmarks}: the locks are partitioned across multiple daemon managers, which serve lock requests received over the network (using RDMA in MIND). Each daemon manager maintains a wait queue per lock and handles acquisition and release requests similar to any reader-writer lock.

%

\paragraphb{Evaluation setup} We use a cluster with five servers connected via a programmable switch to deploy \name atop MIND~\cite{mind}. The switch has a 32-port 6.4 Tbs Tofino programmable switch ASIC. One of the servers is equipped with two 18-core Intel Xeon processors, 384GB memory, and four Mellanox CX-5 100 Gbs NICs, and is used to host a single memory blade VM. The remaining four servers are equipped with two 12-core Intel Xeon processors and two Mellanox CX-5 100 Gbs NICs each and host two compuante blade VMs per server (one per socket), each with 512MB DRAM and 10 cores (with the remaining $2$ cores dedicated to the OS). 
MIND supports the transparent execution of multi-threaded shared memory applications; \name preserves this transparency.

\paragraphb{Real-world applications and workloads} We consider two applications: an in-memory key-value store (dubbed MIND-KVS) from MIND~\cite{mind} that supports fine-grained locks, and Kyoto Cabinet~\cite{kyotocabinet} that supports coarse-grained locks. MIND-KVS employs a hash table where each hash bucket is protected by a fine-grained reader-writer lock. For \name, we port it to combine the first 4 KB of data in each hash bucket with the lock by using \name's Rust API~(\S\ref{ssec:designapi}). We evaluate it with YCSB workloads A, B, and C~\cite{ycsb_workload}, corresponding to $50\%-50\%$, $95\%-5\%$ and $100\%-0\%$ reader-writer proportions, respectively. Kyoto Cabinet is a database management system that employs a database-wide global lock, commonly used as a benchmark in prior works~\cite{cloflock, cnalock, lockelision, vsync}. We run it with the TPC-C workload~\cite{tpcc_workload} with high and low contention, \ie,~1 and 10 warehouses, respectively. All TPC-C transactions hold the global lock in exclusive mode to ensure atomicity and use \name's \code{Pthread} API without the combined data optimization~(\S\ref{ssec:designoptimizations}). For both applications, we run $10$ concurrent worker threads on each compute blade (one worker per core) that continuously generate client requests from the YCSB and TPC-C workloads.

\begin{figure*}[ht!]
\centering
    \includegraphics[width=0.9\textwidth]{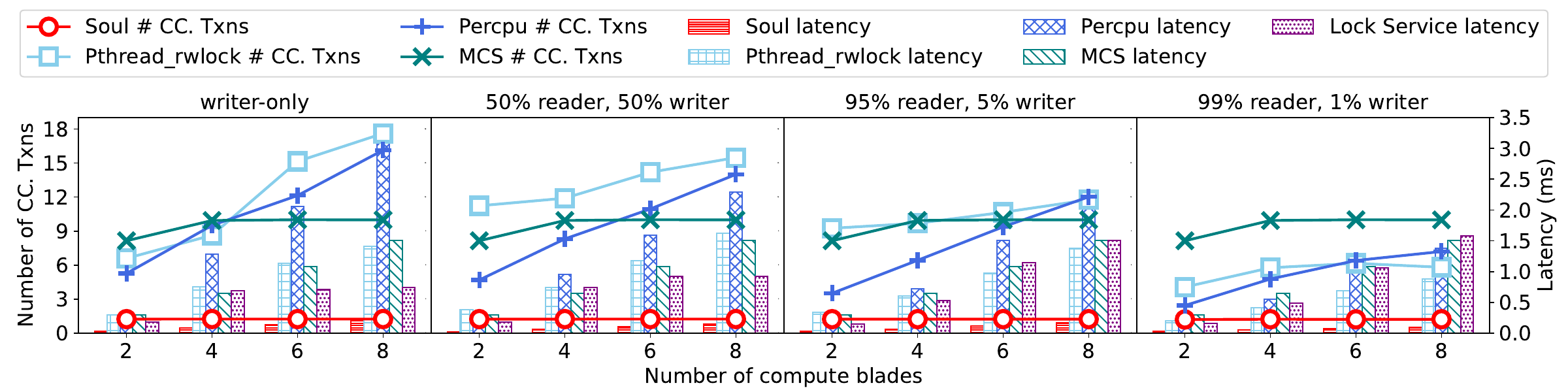}
  \vspace{-10pt}
  \caption{\textbf{Average latency and number of cache-coherence transactions per lock and acquisition~(\S\ref{ssec:understandperf}).}
  \name incurs one cache-coherence transaction per lock and data acquisition, unlike the other systems where cache-coherence transactions scale linearly with the number of blades (\texttt{Pthread} and \texttt{Percpu}) or remain at a large constant (\texttt{MCS}) due to the inefficiency of layering atop cache coherence.
  }
  \label{fig:eval_micro_perf}
  \vspace{-1em}
\end{figure*}

\subsection{Performance for Real-World Workloads}
\label{ssec:eval_macro}

\paragraphb{MIND-KVS} Fig.~\ref{fig:eval_kvs} shows that \name outperforms the compared systems across various reader-to-writer ratios. In particular, \name performs better as the ratio of readers increases, enabling linear scaling for YCSB-C (read-only) --- \name achieves $37.1$ Mops at $8$ blades, $2$-$3$ orders of magnitudes higher throughput compared to \texttt{Pthread}, \texttt{Cohort}, \texttt{MCS}, and \texttt{Lock-Service}. This is because those systems write to their lock variables even when acquiring a read lock, resulting in heavy cache invalidations over the network. Similarly, \texttt{Lock-Service} also requires application threads to send acquisition requests over the network to the corresponding manager daemon for reader locks. On the other hand, \name does not require cache invalidations when there are no writers, as it exploits temporal locality outlined in \S\ref{ssec:designoptimizations}. As such, the most frequently accessed locks and data in YCSB can concurrently remain cached across multiple compute blades.

Although \texttt{Percpu} shows similar performance to \name for the read-only workload (YCSB-C) due to its fully distributed per-core-reader-indicator, it faces significant performance degradation even with $5\%$ of writers (YCSB-B) due to the inefficiency of inter-blade communication through a layered design. \texttt{Lock-Service} observes significantly lower throughput than others at a small number of blades or high reader ratios since it does not leverage the temporal locality of lock accesses, \ie, every lock acquisition must go over the network, unlike \name and other lock algorithms. 
\paragraphb{Kyoto Cabinet} Fig.~\ref{fig:eval_kc} shows that \name's performance is comparable to the best among the compared systems. Since Kyoto Cabinet's global lock design allows only one transaction to be executed at a given point in time (unlike the fined-grained lock in MIND-KVS), Kyoto Cabinet's throughput unavoidably decreases with the number of blades due to increased contention. \name and \texttt{Cohort} outperform other approaches due to their memory hierarchy-aware design.

\paragraphc{Key takeaway:} \name observes better absolute performance as well as performance scaling relative to locks layered atop cache-coherence substrates and software-based standalone lock services, due to a combination of reduced cache-coherence transactions in \designname (\S\ref{ssec:generalizedcc}) and optimizations adapted from traditional cache-coherence protocols (\S\ref{ssec:designoptimizations}).

\begin{figure*}[ht!]
\centering
    \includegraphics[width=0.92\textwidth]{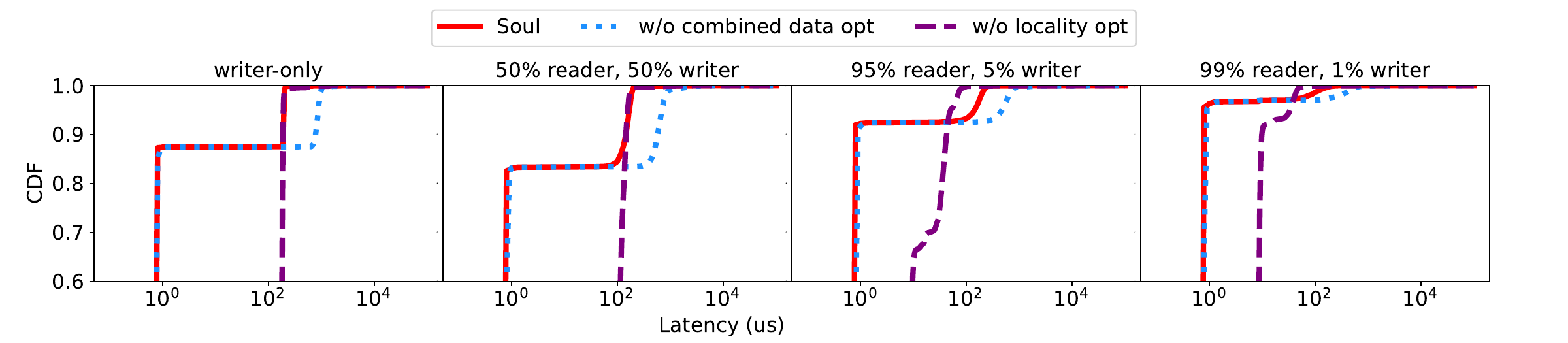}
  \vspace{-10pt}
  \caption{\textbf{Latency CDF of lock and data acquisition for \name with and without the optimizations listed in~(\S\ref{ssec:designoptimizations}).} The locality optimization reduces much of the acquisition latency by caching the lock and shared data. The combined data optimization further reduces acquisition latency by eliminating a network round-trip for data retrieval. Note that the x-axis is in log-scale.}
  \label{fig:eval_micro_understand_opt}
  \vspace{-1em}
\end{figure*}

\subsection{Understanding \name's Performance}
\label{ssec:understandperf}

We further investigate the inefficiency stemming from layering locks atop cache coherence by subjecting all compared schemes to various levels of locking contentions and reader-to-writer ratios. We deploy a thread on each compute blade to contend on a single lock; each thread repeatedly acquires the lock, accesses the shared data ($4$ KB), and releases the lock. We omit \texttt{Cohort} since memory hierarchy-aware locks are identical to their underlying lock for inter-blade synchronization (\eg, \texttt{Pthread} lock in our evaluation). We measure the average latency of inter-blade lock and data acquisition (right y-axis in Fig.~\ref{fig:eval_micro_perf}) and the number of cache-coherence transactions per lock and data acquisition (left y-axis in Fig.~\ref{fig:eval_micro_perf}) to highlight the inefficiency incurred by layering. Since \texttt{Lock-Service} does not use cache coherence for locks, we do not report cache-coherence transactions for it.

By design, \name always triggers a single cache-coherence transaction per lock and data acquisition regardless of workload, which is the minimal possible. In contrast, other lock algorithms either trigger cache-coherence transactions proportional to the number of blades (\texttt{Pthread} and \texttt{Percpu}), or trigger a constant but large number of cache-coherence transactions (\texttt{MCS}), due to the inefficiency of building atop the cache-coherence substrate (\S\ref{ssec:layereddesign}).
%
As a result, \name observes $100$-$200~\mu{s}$ lock and data acquisition latency on average at $8$ blades across all workloads, which is one order of magnitude lower than the fastest compared lock algorithm in each workload. Although \texttt{Lock-Service} is not built atop the cache-coherence substrate, it still observes a similar latency as the other compared systems due to the additional network latency for fetching lock and data separately and the software latency for processing lock requests at the manager threads.


\subsection{Contributions of \name's Optimizations}
\label{ssec:understandopt}

We break down the contributions of \name's optimizations (\S\ref{ssec:designoptimizations}) by comparing the latency distribution of lock acquisition and data fetch with the same setup in \S\ref{ssec:understandperf} for three schemes: (1) \name with all optimizations enabled, (2) \name without the optimization for combining lock acquisition with data fetch (dubbed \code{w/o combined data opt} in figures), and, (3) \name without the optimization that leverages temporal locality of locks and associated data (dubbed \code{w/o locality opt}). 
For scheme (2), the data fetch is triggered by MIND's cache-coherence protocol, while for scheme (3), the lock and its associated data are evicted once the lock is released. 

As shown by the gap between \name and \code{w/o locality opt} in Fig.~\ref{fig:eval_micro_understand_opt}, \name's locality optimization reduces much of the acquisition latency --- from tens or hundreds of microseconds to under 1 microsecond --- by caching the lock and shared data, decreasing the need for expensive network communications. The gap between \name and \code{w/o combined data opt} illustrates how \name's combined data optimization reduces acquisition latency by eliminating the need for one additional network round-trip for data retrieval.

\yanpengdelete{
For writer locks (Fig.~\ref{fig:eval_micro_w_inter} and~\ref{fig:eval_micro_w_inter_lat}), the throughput decreases sharply from one blade to two blades, both with and without the combined data fetch optimization. As described in \S\ref{ssec:eval_macro}, this is because two or more compute blades necessitate network communications due to invalidations from other writers. As the number of blades increases from 2 to 8, \name maintains a constant writer lock throughput ($\sim 0.3$ Mops) regardless of contention, with a linearly increasing latency. Again, as noted in \S\ref{ssec:eval_macro}, this is the ideal performance for any reader-writer lock, since only one writer can hold a lock at any point in time --- as the number of writers (blades) increases, each waiter must wait longer, with a constant throughput across all writers. Even without the combined data and lock acquisition optimization, \name maintains the constant throughput and linearly increasing latency. However, there is a $\sim 6\times$ reduction in throughput ($\sim 0.05$ Mops) and increase in latency relative to \name with the optimization, since lock and data fetch must now occur over two sequential cache-coherence transactions. Note that locality optimization has no effect on writer-writer contentions since the lock and data are forcibly invalidated over the network by another writer (on a different blade) after every acquisition. Finally, we note that the absolute latency for acquiring a writer lock is significantly higher than that for acquiring a reader lock; this is primarily because, unlike reader locks, acquiring a writer lock requires the thread to wait in the queue until the previous writer explicitly releases the lock. Combined with additional network delays for lock acquisition and queue transfers, this results in higher latencies.

\paragraphc{Key takeaway:} For the inter-blade setup, \name's locality optimization significantly improves throughput, latency, and performance scaling for reader-reader contentions, while its combined data/lock acquisition optimization does the same for writer-writer contentions.

\paragraphb{Intra-blade scaling} As expected, the throughput for compared schemes scales linearly with the number of threads since each thread contends for a separate lock (Fig.~\ref{fig:eval_micro_r_intra} and~\ref{fig:eval_micro_r_intra_lat}). Also, the performance trends across different compared schemes are nearly identical to those for inter-blade scaling --- our locality optimization permits near-linear scaling with the number of threads per blade, while \name without the optimization observes lower throughput, higher latency, and worse scaling since they must issue requests over the network for every lock acquisition and release.

Similar to reader locks, while writer lock throughput increases linearly with the number of threads per compute blade, the lock acquisition latency also increases slightly (Fig.~\ref{fig:eval_micro_w_intra} and~\ref{fig:eval_micro_w_intra_lat}); we found the root cause behind this to be increased queueing at the RDMA NIC processing units (PUs); similar observations of RDMA NIC PU saturation at high request rates and large requests have been reported in prior work~\cite{storm_systor_19}. We believe future RDMA NICs with more efficient PUs may eliminate these bottlenecks. Again, the trends across schemes remain similar to inter-blade scaling: the combined lock/data acquisition significantly improves both absolute performance ($3.7\times$ to $6.2\times$ higher throughput, $71\%-85\%$ lower average latency) and performance scaling by reducing contention.

\paragraphc{Key takeaway:} The intra-blade setup observes similar benefits from \name's locality and combined data/lock acquisition optimizations. While reader performance scales linearly with the number of threads per blade, writer performance scales sub-linearly due to RDMA NIC PU saturation.
}

\section{Limitations and Future Research}
\label{sec:futurework}

We outline key limitations of generalized cache coherence and \name, along with future research directions they expose.

\paragraphb{Generalizing other cache-coherence protocols} While both our generalized cache coherence and the \name implementation of it work with directory-based MSI protocol, our design is still compatible with more complex cache-coherence protocols like MESI~\cite{mesi}, MESIF~\cite{mesif}, MOSI~\cite{mosi}, and MOESI~\cite{moesi}. These protocols typically enable further scalability improvements by introducing additional permissions for reducing coherence traffic triggered by common-case coherence transactions. They also, however, introduce more intermediate states and more complex coherence transactions, requiring not only more careful adaptation of transactions to our generalized protocol but also more resources (\eg, directory state and logic) for realizing a feasible implementation for \name.
We leave their generalization to future work.

\begin{figure}[t!]
\centering
\includegraphics[width=0.45\textwidth]{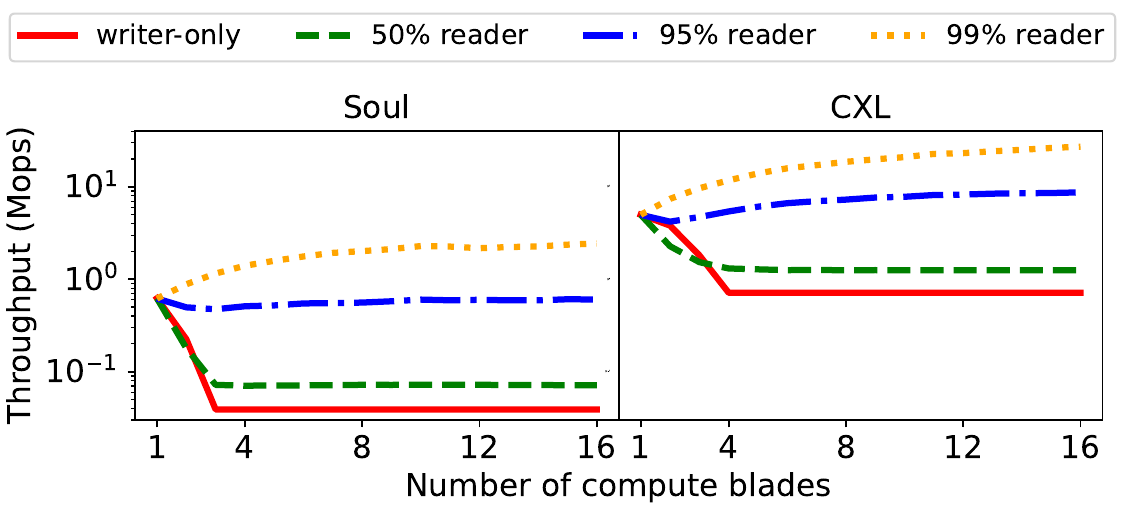}
\vspace{-10pt}
\caption{\textbf{Simulated performance scaling for \name and \designname with CXL performance.} The latter delivers an order of magnitude higher throughput than \name and enables near-linear scaling for read-mostly workloads~(\ie,$99\%$ reader) under contention.}
\label{fig:futurework_cxl}\vspace{-1em}
\end{figure}

\paragraphb{Generalized cache coherence on other platforms} \name implements generalized cache coherence on a specific architecture for directory-based MSI coherence; many other variants of the architecture exist, \eg, those with distributed (partitioned) directories placed closer to compute units or memory units. Moreover, our implementation focuses on a cache coherent substrate that leverages Ethernet and programmable switches; emerging high-performance interconnects like Compute eXpress Link 3.0 (CXL 3.0)~\cite{cxl} leverage higher throughput and lower latency PCIe, and place the directory at the memory devices for rack-scale memory pooling across heterogeneous compute devices. In order to understand the performance benefits \designname can bring to CXL, we studied how an implementation would perform at scale. Since CXL 3.0 hardware is not yet commercially available, we built a \designname-enabled disaggregated memory simulator that can assume either Ethernet or CXL performance characteristics~\cite{pond}. Specifically, the simulator implements a disaggregated memory system that deploys a directory-based MSI substrate with cache directories maintained at the memory device, with the same wait queue (\S\ref{ssec:implwaitqueue}) and shared memory list (\S\ref{ssec:implmultilocationentry}) implementation as \name. We evaluate the simulator under the same settings as \S\ref{ssec:understandperf} and measure throughput scaling with the number of compute blades. Fig.~\ref{fig:futurework_cxl} shows that compared to \name, a \designname operating at CXL performance will not only delivers $1$ orders of magnitude higher synchronization throughput but also enables linear scaling for read-mostly (\ie,$99\%$ reader) workloads even under contention due to lower inter-compute or compute-memory latency ($\approx300\mu{s}$). Exploring a hardware realization of this approach for CXL and other architectures offers exciting future research opportunities.


\section{Related Work}
\label{sec:background}
We already discussed works related to \designname and \name in \S\ref{sec:background}; we now discuss a few additional related approaches.


\paragraphb{Lock-based synchronization} Locks are the most widely used synchronization primitive in shared memory programming and are critical to parallel programs' performance and scalability~\cite{nonscalablelocks, litl, everthingsync,mcslock, clhlock, cohortlock, hmcslock, cnalock, shufflelock, cloflock, linuxbrlock, passiverwlock, numarwlock, bravo}.  Queue-based locks~\cite{mcslock, clhlock} achieve constant cache-coherence cost on multi-core machines regardless of contention by ordering all waiting requestors into a queue and limiting direct communication between adjacent requestors in the queue. On NUMA architectures, memory hierarchy-aware locks further optimize cache-coherence traffic by prioritizing intra-NUMA node communication over inter-NUMA node communication~\cite{cohortlock, hmcslock, cloflock,cnalock, shufflelock}. \name leverages techniques from both queue-based and hierarchical locks.  

\section{Conclusion}



In this work, we have argued for a co-design of cache coherence and synchronization. Our driving observation is that lock-based synchronization is essentially a generalization of cache coherence in time and space. We incorporate this insight into \designname, a novel class of \textit{G}eneralized \textit{C}ache-coherence \textit{P}rotocols for lock-based synchronization, and demonstrate \name as an implementation of \designname atop disaggregated memory. Our evaluations show that \name improves in-memory key-value store and database management system performance at scale by $1-2$ orders of magnitude.

\section*{Acknowledgments}
This work is supported in part by NSF Awards \#2112562, 2147946, 2118851, and 2047220, as well as a NetApp Faculty Fellowship.

\newpage
{
\balance
\bibliographystyle{lowerCaseTitle}
\bibliography{bib/abr-short,bib/paper}
}
\begin{sloppypar}


\end{sloppypar}

\end{document}